# Effect of emotions and personalisation on cancer website reuse intentions

Sunčica Hadžidedić [a, b], Alexandra I. Cristea [a, b], Derrick G. Watson [c]

[a] Department of Computer Science, Durham University, Durham, DH1 3LE, United Kingdom
[b] Department of Computer Science, University of Warwick, Coventry, CV4 7AL, United Kingdom
[c] Department of Psychology, University of Warwick, Coventry, CV4 7AL, United Kingdom

**Abstract**: The effect of emotions and personalisation on continuance use intentions in online health services is underexplored. Accordingly, we propose a research model for examining the impact of emotion- and personalisation-based factors on cancer website reuse intentions. We conducted a study using a real-world NGO cancer-support website, which was evaluated by 98 participants via an online questionnaire. Model relations were estimated using the PLS-SEM method. Our findings indicated that pre-use emotions did not significantly influence perceived personalisation. However, satisfaction with personalisation, and perceived usefulness mediated by satisfaction, increased reuse intentions. In addition, post-use positive emotions potentially influenced reuse intentions. Our paper, therefore, illustrates the applicability of theory regarding continuance use intentions to cancer-support websites and highlights the importance of personalisation for these purposes.

## 1. Introduction

The demand for digital health has been rising [1, 2]. However, challenges and barriers to its usage have also become more evident. A number of prior studies, thus, advocated for personalisation [3], i.e., tailoring content to user needs, to improve engagement with digital health systems [2], and increase user satisfaction and usage intentions [4].

Online health service personalisation has been explored from the perspective of: its association with self-disclosure [5], systematic literature reviews which have highlighted its benefits [6], personalised educational health content for the elderly [7] and adolescents [8], distress-based personalised therapy recommendations [9], and recommending health and fitness content for runners [10]. Nevertheless, research exploring the different factors that might influence intentions to (re)use personalised online health services is lacking. Our study addresses this gap, specifically for cancer-support websites.

Moreover, much has been discovered about how information technology (IT) use intentions are affected by cognitive factors, as are perceived usefulness and satisfaction [11-15]. However, there is insufficient understanding of how a user's affective states might influence their use of online personalisation [16, 17], although it is known that emotions impact human behaviour and perception [18], and are induced by different interactions [19]. Therefore, there is good reason to explore if emotions can lead to and/or be influenced by the use of personalised online health services.

Indeed, persistent sadness or anxiety were shown to increase the likelihood of online health information seeking [20]. Some studies examined the effects of website design [21], interface aesthetics [22] and website features [23] on affective responses (e.g., arousal and irritation). Furthermore, the possible links between emotions and online personalisation have been considered in the domains of: e-commerce [4], group decision support systems [24], e-learning [17] and emotion-aware recommender systems [9, 25, 26]. However, to the best of our knowledge, our study is the first to propose a research model that *explores emotions and perception of personalisation as factors influencing continuance use intentions* in the context of cancer-support websites.

## 2. Theoretical background and research model

### *2.1. Underlying theories*

The underlying theories for our conceptual framework were: i) the *two-stage model of cognition change toward IT usage* [27] (Figure 1, part 1), ii) *affect theory* (Figure 1, part 2), which defines emotions as drivers of human behaviour [18], iii) and *appraisal theories* (Figure 1, part 3), which describe emotions as reactions [19] to the current context's assessment [28, 29]. The two-stage model has previously been applied to, e.g., digital learning technologies [15], however, not to personalised cancer-support websites. The model measures changes in beliefs and attitude from pre- to post-usage stages, and satisfaction as a post-usage affective state.

Our conceptual framework integrates the belief- and affect-based constructs at different IT use stages, however, we only examine changes in affective state. Furthermore, we measure belief only at the post-use stage, specifically as the perceived usefulness of personalisation. In the two-stage model, the satisfaction construct captures the extent of user satisfaction, pleasantness, content, and delight toward IT. We, on the other hand, use this construct to represent post-use cognitive-based appraisal of different aspects of personalisation (Section 2.2.4 and Table A1).

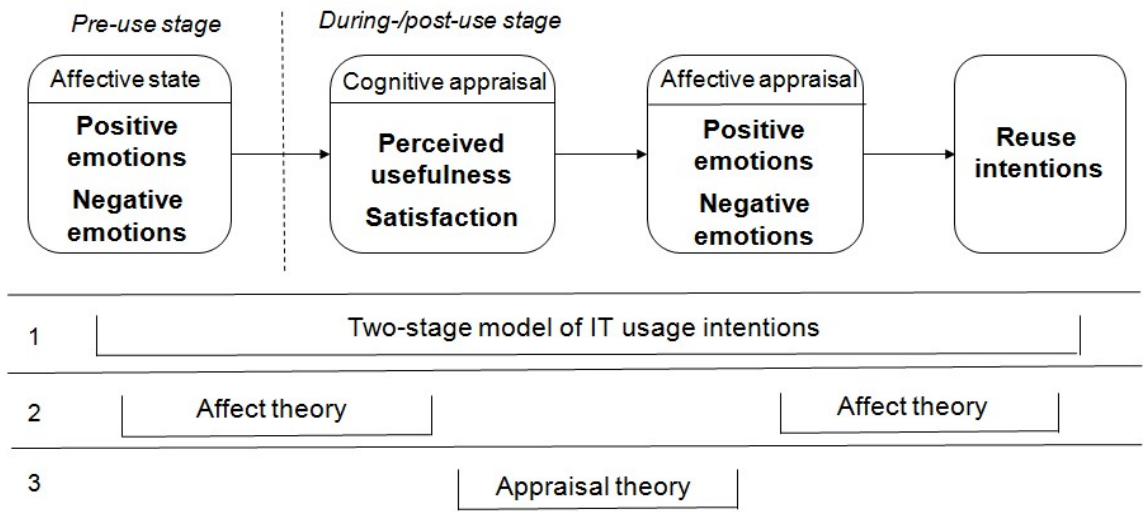

**Figure 1.** Conceptual framework - expanded two-stage model of IT usage intentions with emotions

Importantly, the two-stage model provides only a limited understanding of the relationship between emotions and perceptions or behavioural intentions, i.e., via satisfaction. Pre- and post-usage attitudes measure user perception (i.e., cognitive appraisal) of whether system use was good, wise, positive and effective. In contrast, our pre- and post-use emotions' factors explicitly capture user emotions, which are a very specific affect type (Section 2.2.2). Hence, our framework extends the two-stage model. It incorporates emotion-specific constructs, which were derived from affect theories [18, 19] and measured by user self-assessment of *basic emotions'* intensities experienced at a particular moment of system use.

## 2.2. Research model and hypotheses

Building upon the conceptual framework from Section 2.1, Figure 2 illustrates our research model. This section defines its constituting factors and hypotheses.

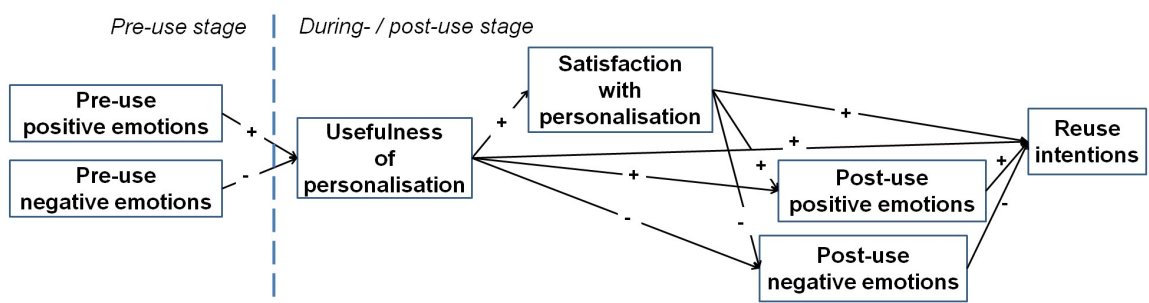

**Figure 2.** Research model

*2.2.1. Reuse intentions*

Behavioural intention is an intent to perform certain behaviour [1]. Theories of reasoned action and planned behaviour introduced this concept, followed by its adoption in IT use [30] and continuance intentions [27] frameworks, and applications in, e.g., online personalisation [4, 31] and online health services [1, 32, 33]. Our study measures users' intentions to revisit and use a personalised cancer-support website.

*2.2.2. Emotions (pre- and post-use)*

Emotions are high intensity, brief affective states [24] that begin quickly [34]. We measured 12 emotions, selected based on the results from two prior studies. The first study [35] explored the likelihood that interest, fear, sadness, surprise, awe, anger, embarrassment, guilt, enjoyment, shame, happiness, contempt, or disgust stimulate online cancer information searching. The second study [36] examined the association between online personalisation needs, usage intentions, and 13 basic and possible basic emotions, as classified by [34, 37].

These two studies showed that only certain basic emotions play a significant role in online health information use or perceived personalisation needs. These were fear, sadness, *awe, excitement, interest* and *surprise*, and were hence all accounted for in our study. We also added six other emotions. Anxiety, boredom and *calmness* (or neutral state) were included due to their frequent use in the related human-computer interaction (HCI) research [16, 38-41]. Embarrassment, guilt and *happiness* were basic emotions that were re-evaluated in this study to balance the number of positive and negative emotions measured. Happiness (or alternatively joy, enjoyment, pleasantness) is also one of the essential positive valence emotions often studied in HCI [42]. Based on vulnerability research [43], embarrassment and guilt were included to represent the cancer-affected people's potential negative perception of their own self, state or actions.

Given that positive and negative emotions influence behaviour differently [18], we used previous research [29, 34, 44-47] to classify the 12 emotions into positive (denoted above in *italics*) and negative valence categories. Furthermore, we measured emotions at two stages (Section 2.1): pre-use emotions at the beginning of website use, and post-use emotions after website use.

*Emotions as stimuli of perception or action*

The effect of emotions on IT perception has been addressed in a few studies. For example, emotional attachment influenced the perceived usefulness and attitude towards Facebook [48], and affective quality of smart watches was associated with their perceived usefulness [13]. However, such research in the online health domain is very limited, e.g., indicating that interest and excitement increased the perceived needs for personalisation [36]. Given the under-researched association between emotions and perceived personalisation, we hypothesised that:

*H1.1: Pre-use positive emotions increase the perceived usefulness of personalisation.*
*H1.2: Pre-use negative emotions decrease the perceived usefulness of personalisation.*

Interestingly, there is more research on the association between emotions and behavioural intentions towards IT. Enjoyment was found to influence web use [41], anxiety influenced continuance use intention for mobile-health among elderly [49], emotions stimulated online search for cancer information [35], and interest increased cancer patients' use of electronic health records [50]. Strong positive emotions, or absence of negative emotions, mediated the effect of personalisation on online purchase intentions [51]. However, we found only one prior study [36] that used linear regression and showed a positive influence of interest on reuse intentions of partially personalised online cancer services.

Due to the lack of research on cause-effect relations (or structural equation modelling) between emotions and user intentions on personalised cancer-support websites we hypothesised that:

*H2.1: Post-use positive emotions increase reuse intentions.*
*H2.2: Post-use negative emotions decrease reuse intentions.*

*Emotions as reactions or affective appraisal*

We argue that users' appraisal of cancer-support website personalisation can evoke post-use positive and negative emotions. This is based on research showing that certain website features (e.g., colour, images, shapes) induced emotions [52], perceived usefulness of educational blogs increased liking and pleasantness [11], pedagogical agent's adaptation intensified enjoyment and decreased boredom [42], personalisation predicted post-use positive emotions in online shopping [51], and online health information overload (i.e., lacking personalisation) influenced negative emotions [53]. However, there is insufficient understanding about the impact of personalised online health services [52] on affective states. We therefore postulated the following hypotheses:

*H3.1: Perceived usefulness of personalisation positively influences post-use positive emotions.*
*H3.2: Perceived usefulness of personalisation negatively influences post-use negative emotions.*

Moreover, consistent with research that showed that satisfaction positively influenced attitude after using IT [27, 54], we hypothesised that:

*H4.1: Satisfaction with personalisation positively influences post-use positive emotions.*
*H4.2: Satisfaction with personalisation negative influences post-use negative emotions.*

*2.2.3. Usefulness of personalisation*

Perceived usefulness relates to expectations about performance improvements as a result of using a service or product [54]. Our *usefulness of personalisation* factor adapted Davis's [55] 'perceived usefulness' to evaluate the individual personalisation features implemented on the studied cancer-support website. This approach, has also been applied to e-commerce [44] and personalised e-learning [56].

Previous work has shown that perceived usefulness had a significant positive impact on satisfaction in the use of, e.g., online- and m-banking [57, 58], e-government [54], and digital textbooks [15]. Moreover, personalisation applied to online news [59], e-commerce [60] and online-banking [61] increased user satisfaction. Therefore, for cancer-support websites, we hypothesised that:

*H5: Perceived usefulness of personalisation increases satisfaction with personalisation.*

Furthermore, previous research has argued that perceived usefulness [55] is an essential criterion for system reuse [62]. It increased usage intentions for digital textbooks [15], health-information portals [32] and mobile health applications [63]. Hence, we hypothesised that:

*H6: Perceived usefulness of personalisation increases reuse intentions.*

*2.2.4. Satisfaction with personalisation*

In a seminal work on customer expectations, Oliver [64] defines satisfaction as the "summary psychological state resulting when the emotion surrounding disconfirmed expectations is coupled with the consumer's prior feelings about the consumption experience". Abundant research shows that satisfaction positively affects online repurchase intentions [65] and (continuance) usage intentions for online-banking [57], e-government [54], educational blogs [11], m-health [33], and personalised information reuse [60]. Accordingly, with respect to personalised cancer-support websites, we hypothesised that:

*H7: Satisfaction with personalisation increases reuse intentions.*

## 3. Methodology

### 3.1. *Study design and data collection*

We sampled people directly and indirectly affected by cancer, i.e.: former/current patients; caregivers – family and friends; and those preventatively seeking cancer information. This was achieved by using:
- purposive sampling for cancer association members;
- and convenience sampling for university students (as primary users of online health services [66]), social networks' users, and crowdsourced participants (via Amazon Mechanical Turk[1]).

This study was reviewed and approved (REGO-2015-1421) by the Biomedical and Scientific Research Ethics Committee at the University of Warwick. It used an online survey methodology. The survey first explained the study's objective, the right to withdraw without consequences, and informed participants they were consenting to take part in this research and the collection of their anonymised responses. The following were then presented in a consecutive order:
(1) *Questionnaire*: Reporting pre-use emotions (≈ 5 minutes).
(2) *Experiment:* User-website interaction (≈ 25 minutes).
(3) *Questionnaire*: Post-use emotions and website evaluation (≈ 30 minutes).

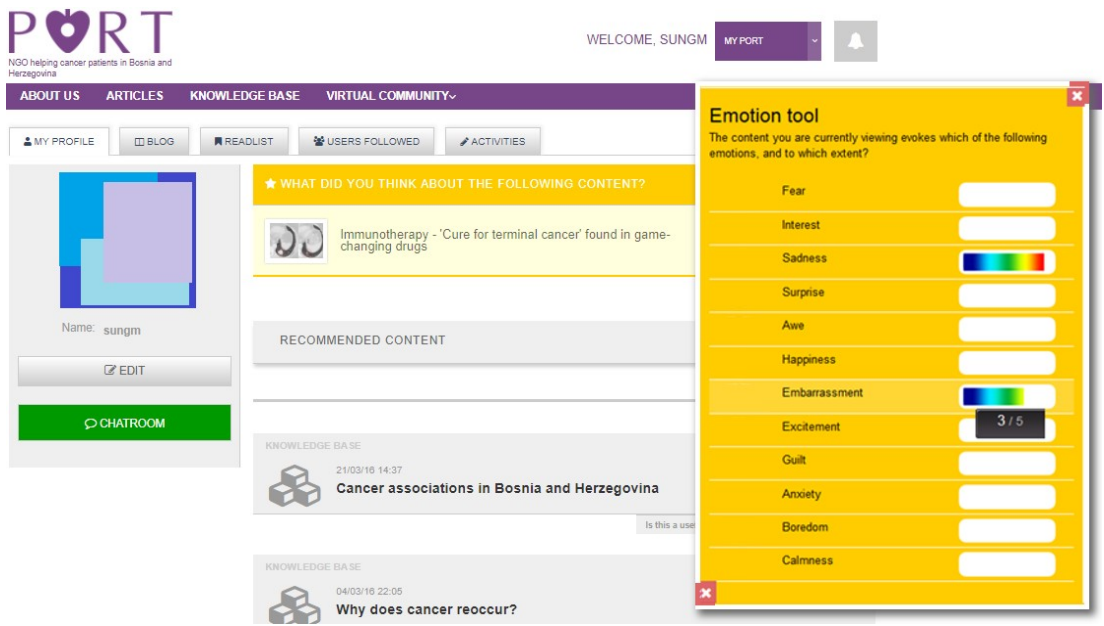

**Figure 3.** PORT cancer website

---

[1] https://www.mturk.com/

During the experiment (step 2), participants interacted with PORT [67], a personalised cancer-support website (Figure 3) for patients and caregivers. PORT's cancer-related content included cancer patients' blogs, and articles adopted from respectable online sources about different cancer types, treatments and therapies. Participants completed the following[2]: user-profile creation; privacy policy customisation; user-profile editing; interface adaptation (e.g., adjusting font, colour, etc.); rating content and reviewing recommendations. Since we were interested in the effect of interaction with a personalised cancer-support website, these tasks were essential for a user to explore the website, receive and perceive the personalisation, which on PORT comprised cancer information recommendations and user interface adaptation. The questionnaires (steps 1 and 3; see Appendix A) collected data on pre-use emotions, user demographics, post-use emotions, perceived usefulness of and satisfaction with personalisation, and website reuse intentions.

The scale for measuring emotion intensity was adopted from a game experience questionnaire [68], applied to online systems [44, 69]. Items from validated instruments were used for *satisfaction with personalisation* [70, 71] and *reuse intentions* [51, 72]. The *perceived usefulness* instrument [55] was modified to measure the usefulness of individual personalisation features implemented on the PORT website (see Appendix A).

The online survey started in May 2015, and ran for 1.5 months. We received 122 responses; 98 were valid and used in data analysis. We removed the data from respondents who were not affected at all or not interested in cancer.

### 3.2. Data analysis and instrument validation

Data pre-processing, exploratory factor analysis (EFA) and descriptive analyses were conducted using IBM SPSS® Statistics[3]. SmartPLS 3[4] was used for confirmatory factor analysis (CFA) and structural equation modelling (SEM) with partial least squares (PLS) method.

EFA was only applied to the 24 items for *usefulness of personalisation* (Appendix A), as we modified the original instrument. We used principal axis factoring [73], direct Oblimin rotation [74], with Kaiser normalisation and a fixed number of factors based on our previous studies, which were confirmed with Eigenvalues>1.0 and a scree test [73]. A two-factor solution was selected: 55.89% variance explained; Eigenvalues>1.43; KMO = 0.76; $\chi2(55)$ = 440.03, $p$ < 0.001. 7 items reflected the factor *usefulness of content-related personalisation* (UsfCP), and 4 items represented the factor *usefulness of explicit UI- and*

---

[2] See Appendix A for the complete list of website features participants were exposed to, asked to interact with and evaluate on perceived usefulness.

[3] https://www.ibm.com/uk-en/products/spss-statistics

[4] https://www.smartpls.com/

*content-adaptation* (UsfADP). Namely, UsfCP covers the automatically generated recommendations of different content (e.g., articles, blog posts) and the content rating functionality required for these purposes. UsfADP, on the other hand, refers to website features requiring more explicit user involvement for content customisation (e.g., notifications and privacy policy length) and text appearance adaptation.

We next ran a CFA in SmartPLS on the refined model with eight factors: pre-use positive (PREPE) and negative emotions (PRENE), post-use positive (POPE) and negative emotions (PONE), usefulness of content-related personalisation (UsfCP), usefulness of UI-/content-adaptation (UsfADP), satisfaction (SAT) and reuse intentions (RI). Cronbach's α and composite reliability ≥0.7 [75] and AVE - average variance extracted >0.5 [76] were achieved by iteratively removing items with low outer loadings - starting with <0.5, up to 0.7 [73, 77]. Table 1 presents reliability and validity results, and Table A1 (Appendix A) factor loadings. The Fornell-Larcker criterion for discriminant validity was satisfactory [78]. Heterotrait-monotrait ratio [79] was <0.85 for all factors; apart from *pre-use* to *post-use negative emotions* (HTMT = 0.907), likely due to the same constituting emotions: fear, sadness and guilt/embarrassment). However, the latter was acceptable at the HTMT$_{inference}$ criterion [79].

**Table 1.** Construct reliability and validity

| Factor | Num. of items | Mean (SD); N | Cronbach's α | Composite Reliability | AVE |
|---|---|---|---|---|---|
| PREPE | 2 | 1.6 (0.8); 98 | 0.731 | 0.734 | 0.581 |
| PRENE | 3 | 1.6 (0.8); 98 | 0.771 | 0.771 | 0.529 |
| UsfCP | 4 | 3.9 (0.7); 97 | 0.826 | 0.828 | 0.546 |
| UsfADP | 3 | 3.8 (0.8); 97 | 0.757 | 0.754 | 0.507 |
| SAT | 3 | 3.9 (0.7); 96 | 0.797 | 0.796 | 0.566 |
| POPE | 2 | 1.8 (0.9); 98 | 0.734 | 0.734 | 0.580 |
| PONE | 3 | 1.7 (0.8); 98 | 0.761 | 0.763 | 0.520 |
| RI | 3 | 3.7 (0.8); 98 | 0.820 | 0.820 | 0.604 |

## 4. Results

### 4.1. Participant demographics

The respondents' age ranged from 18 to 57 years (Mean=27, SD=8.9). The majority were from B&H (51%) and USA (33.7%), and 61.2% were female. They were mainly caregivers to a family member who suffered from cancer (54.1%), preventatively sought cancer information (30.6%), had a friend suffering from cancer (14.3%), or were a cancer patient (1%).

## *4.2. PLS-SEM results*

Model fit was tested with a consistent PLS algorithm - all LVs connected for initial calculation, 300 iterations, path weighting scheme, missing values replaced with a mean. SRMR (0.069 *saturated, 0.181 **estimated model) met the recommended value of <0.08 [80], while NFI (0.717*, 0.587**) was slightly below the recommended 0.9-1 [81]. Figure 4 shows the path coefficients (*ß*) and coefficients of determination ($R^2$) after applying complete bootstrapping with 2000 subsamples.

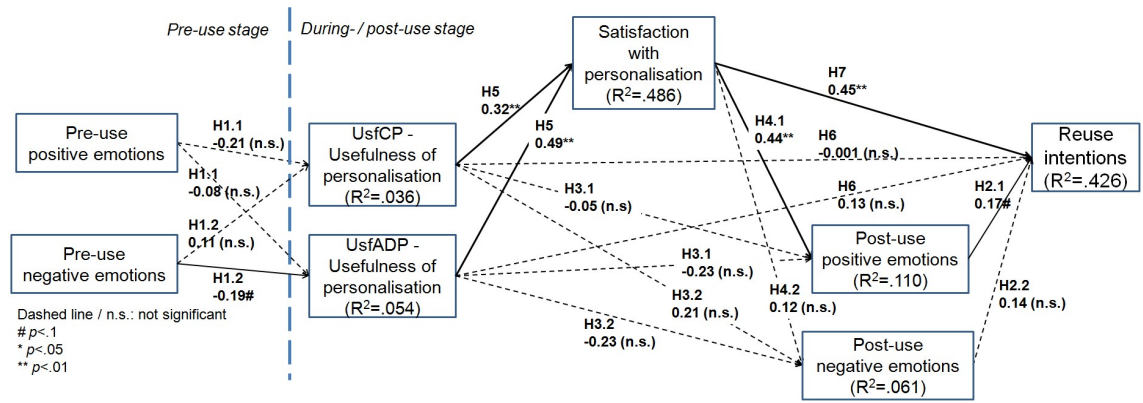

**Figure 4.** Estimated model - path coefficients and significance levels

The findings showed that four path coefficients were significant at *p*<0.05, supporting H4.1, H5, H7. At the pre-use stage, negative emotions (specifically fear, guilt and sadness categories) decreased the usefulness of adaptation-related personalisation, however at *p*<0.1 (H1.2: *ß* = -.19, *t* = 1.71, *p* = .088). During website use, perceived usefulness of content personalisation (H5 - UsfCP: *ß* = .32, *t* = 2.93, *p* = .003) and adaptation (H5 - UsfADP: *ß* = .49, *t* = 4.77, *p* = .000) significantly increased satisfaction. However, without a direct effect on post-use emotions or reuse intentions (H3.1, H3.2, H6 were not supported).

At the during- and post-use stage, satisfaction with personalisation intensified positive emotions (H4.1: *ß* = .44, *t* = 3.2, *p* = .001). Satisfaction (H7: *ß* = .45, *t* = 3.6, *p* = .000), and potentially post-use positive emotions (H2.1: *ß* = .17, *t* = 1.7, *p* = .090), increased reuse intentions. Interestingly, post-use negative emotions did not influence and were not influenced by the factors in our model (H2.2 and H4.2 not supported).

We also tested *mediating effects* (Table 2), based on Zhao's method [82]. Post-use emotions were not significant mediators. Nevertheless, satisfaction fully mediated the effect of usefulness of content- and adaptation-related personalisation (UsfCP and UsfADP, respectively) on post-use positive emotions and reuse intentions.

**Table 2.** Mediating effects

| IV | M | DV | P1: IV->M | P2: M->DV | P3: IV->DV | P1·P2 | Result |
|---|---|---|---|---|---|---|---|
| UsfCP | SAT | POPE | 0.32** | 0.44** | n.s. | 0.14# | Full mediation |
| UsfCP | SAT | RI | 0.32** | 0.45** | n.s. | 0.14* | Full mediation |
| UsfADP | SAT | POPE | 0.49** | 0.44** | n.s. | 0.22** | Full mediation |
| UsfADP | SAT | RI | 0.49** | 0.45** | n.s. | 0.22** | Full mediation |

IV: independent variable, M: mediator, DV: dependent variable.
#$p < 0.1$; *$p < 0.05$; **$p < 0.01$

## 5. Discussion

Our findings imply that the essential factor explaining reuse intentions for cancer-support websites is satisfaction with personalisation. It mediates the effect of usefulness of personalisation, and directly increases reuse intentions, as seen in numerous studies on continuance use intentions in other domains [11, 15, 57]. We next discuss and generalise the key results.

First, pre-use emotions do not significantly affect perceived personalisation. Although prior online-health research indicated a possible effect of positive emotions on personalisation needs [36], our study showed that surprise and awe (i.e., positive-valence, high-arousal emotions) do not influence usefulness of personalisation. Furthermore, we found a marginally significant effect of negative emotions, such that fear, guilt and sadness jointly decrease the usefulness of explicit UI- and content-adaptation (UsfADP), i.e., the type of personalisation which requires explicit user involvement. This likely occurs because people in negative affective states are biased towards negative events/occurrences [83], hence might not perceive the benefits of personalisation. Overall, these are valuable findings, providing an insight into the online cancer-support context, and inviting exploration of alternative emotion taxonomies and their association with perceived personalisation.

Second, contrary to our prediction, usefulness of personalisation does not directly impact post-use emotions. However, usefulness of personalisation (i.e., both content-related personalisation and explicit UI- and content-adaptation) intensifies post-use positive emotions when mediated by satisfaction. Our results are consistent with e-commerce research regarding negative emotions [51], however, there, personalisation increased positive emotions [51]. This difference could stem from the different measurement methods: we observed discrete emotions, while Pappas et al. [51] measured positive or negative mood; we evaluated the perceived usefulness of individual personalisation features, and they examined users' willingness to be provided with personalisation.

Third, cognitive perception of personalisation is more important than its affective appraisal [31]. Almost 50% of variation in satisfaction with personalisation is explained by the personalisation's perceived usefulness. Thus, our findings align with the positive effect found in online banking [57, 58], e-government [54] and digital textbooks [15]. While research has addressed the effect of satisfaction on attitude [27, 54], our study was the first to explore its influence on post-use emotions. Specifically, we found that satisfaction with personalisation intensifies post-use positive emotions, indicating a pleasant surprise after confirming positive or disconfirming negative expectations.

Finally, contrary to the findings of prior research in other domains, reuse intentions for personalised cancer-support websites are not significantly explained by post-use negative emotions or perceived usefulness. Post-use negative emotions and benefits of personalisation affected online purchase intentions in [51], and negative affects, depressive symptoms and trait anger reduced online health information search intentions in [53]. Thus, behavioural intentions are possibly context- or task-dependent, or influenced differently by various affective states. In fact, our findings suggest that post-use surprise and awe could increase cancer website reuse intentions, which aligns with the findings for positive emotions in, e.g., online purchasing [31, 51].

## 6. Conclusion

From a *theoretical* perspective, our research implies that the two-stage model's constructs - usefulness and satisfaction - were applicable to understanding continuance use intentions for personalised cancer-support websites. However, alternative theories, e.g., Theory of Constructed Emotion [84], should be used for investigating the cause-effect between emotions and personalisation.

Unlike the theory-proposed effect [18], emotions in the cancer-support website context were not a significant predictor of perceived personalisation or behavioural intentions. Nevertheless, we confirmed that context appraisal [28, 29], i.e., perceived personalisation did evoke emotions. Furthermore, the frequently reported: i) effect of perceived usefulness on satisfaction with IT, and ii) the influence of satisfaction with IT on its reuse intentions, also prevail in cancer-support websites. Our study's important contribution was measuring perceived usefulness and satisfaction in relation to *personalisation*.

Furthermore, this paper offers *practical implications*. Cancer-support website providers should implement personalisation, particularly content recommendations and interface adaptation. These features increase satisfaction and positive emotions, hence stimulate website reuse.

Our research, however, has *limitations*. Although comparable to computer-use studies [17, 85, 86], our sample size was relatively small. The sampled participants here were mainly people indirectly affected by cancer; future research should focus on cancer patients. Our findings' generalisability is currently limited to cancer-support websites. Moreover, alternative emotion taxonomies could be examined and longitudinal studies for a deeper insight into perceptions of personalisation.

In conclusion, this research uniquely applied affect and IT usage theories. Finally, its main contribution is highlighting the importance of the understudied factors – emotions and personalisation - in forming user intentions toward online cancer-related services.

**Appendix A**

Table A1. Overview of questionnaire items, measurement scales, factors and factor loadings

| Factor | Questionnaire items | Outer loadings |
|---|---|---|
| 5-point scale: *1: Not experiencing this emotion at all, 2: Mildly, 3: Moderately, 4: Very, 5: Experiencing this emotion extremely* | | |
| **Pre-use positive emotions** | Awe | 0.807 |
| | Surprise | 0.715 |
| | Calmness, Excitement, Happiness, Interest (*removed*) | <0.7 |
| **Pre-use negative emotions** | Guilt | 0.738 |
| | Fear | 0.734 |
| | Sadness | 0.709 |
| | Anxiety, Boredom, Embarrassment (*removed*) | <0.7 |
| **Post-use positive emotions** | Surprise | 0.762 |
| | Awe | 0.761 |
| | Calmness, Excitement, Happiness, Interest (*removed*) | <0.7 |
| **Post-use negative emotions** | Embarrassment | 0.814 |
| | Sadness | 0.692 |
| | Fear | 0.647 |
| | Anxiety, Boredom, Guilt (*removed*) | <0.64 |
| 5-point scale: *1: strongly disagree* to *5: strongly agree* | | |
| **Usefulness of …** (*I perceive as useful the personalisation feature…*) **…content-related personalisation** (UsfCP) | UsfCP1. Recommendations for forum discussions | 0.777 |
| | UsfCP2. Recommendations for blogs | 0.750 |
| | UsfCP3. Recommendations for articles/news | 0.723 |
| | UsfCP4. Content rating | 0.704 |
| | UsfCP5. Recommendations for knowledge-base content (*removed*) | <0.7 |
| | UsfCP6. Personal readlist (*removed*) | <0.7 |
| | UsfCP7. Categorising content (popularity, recency, etc.) (*removed*) | <0.7 |
| ***…explicit UI- and content-adaptation*** (UsfADP) | UsfADP1. Privacy policy customisation (long/concise) | 0.749 |
| | UsfADP2. Notifications/reminders | 0.727 |
| | UsfADP3. Text size adaptation | 0.657 |
| | UsfADP4. Text colour adaptation (*removed*) | <0.65 |
| …other evaluated personalisation features (*removed after EFA*) | Tailoring background colour; User-profile customisation; Defining interests; Feedback about personalisation usefulness; "What did you think about this content?"; Emotion tool; Filtering search; Recommendations matching user's interests; Recommendations based on ratings; Recommendations based on user similarity; Filtering recommendations; Customising language; Greetings with username | |
| **Satisfaction with personalisation** (*I am satisfied with how PORT's website was personalised to my needs because it…*) | SAT1. … provided content at the right level of detail | 0.793 |
| | SAT2. … provided valuable content to me | 0.751 |
| | SAT3. … could save me time | 0.710 |
| | SAT4. … knew what I wanted (*removed*) | <0.7 |
| | SAT5. … took into consideration my interests and preferences to make recommendations to me (*removed*) | <0.7 |
| | SAT6. … improved my search performance (*removed*) | <0.7 |
| | SAT7. … provided relevant content to me (*removed*) | <0.7 |
| | SAT8. … provided up-to-date content to me (*removed*) | <0.7 |
| **Reuse intentions** | RI1. Overall, I have a positive attitude toward using the website. | 0.839 |
| | RI2. Given the chance, I intend to use the website again | 0.744 |
| | RI3. I would recommend the website to my friends. | 0.744 |
| | RI4. I intend to use the website frequently. (*removed*) | <0.7 |

**Funding**

This research received no specific grant from any funding agency in the public, commercial, or not-for-profit sectors.

**Declaration of conflicting interests**

None to declare.